\documentclass[12pt]{article}
\usepackage{epsf,latexsym}
\usepackage{amsmath,amssymb,amsthm,cite}
\usepackage{graphicx}

\theoremstyle{definition}                    

\theoremstyle{remark}

\numberwithin{equation}{section}             

\newcommand{\de}{\hbox{\rm{d}}}

\newcommand{\bb}{\begin{eqnarray}}
\newcommand{\ee}{\end{eqnarray}}
\newcommand{\eee}{\nonumber\end{eqnarray}}
\newcommand{\qq}{\quad}

\usepackage[ps,dvips,arc,frame]{xy}

\epsfverbosetrue
\newcommand{\ddf}{\hbox{$^f$\hspace{-0.15cm} $\mathcal{D}$}}

\newcommand{\T}{{\rm tr}}

\newcommand{\pp}[1]{\begin{pmatrix} #1 \end{pmatrix}}

\newcommand{\rxyz}[2]{{\begin{xy} 0;<2mm,0mm>:<0mm,2mm>::0;0,
,(5,-2)*{a}
,(10,-2)*{b}
,(15,-1.8)*{\bar{b}}
,(20,-2)*{c}
,(25,-2)*{d}
,(30,-1.8)*{\bar{d}}
,(35,-2)*{e}
,(40,-2)*{f}
,(45,-1.8)*{\bar{f}}
,(2,-5)*{a}
,(2,-10)*{b}
,(2,-15)*{\bar{b}}
,(2,-20)*{c}
,(2,-25)*{d}
,(2,-30)*{\bar{d}}
,(2,-35)*{e}
,(2,-40)*{f}
,(2,-45)*{\bar{f}}
,(5,-5)*\cir(#1,0){}
,(10,-5)*\cir(#1,0){}
,(15,-5)*\cir(#1,0){}
,(20,-5)*\cir(#1,0){}
,(25,-5)*\cir(#1,0){}
,(30,-5)*\cir(#1,0){}
,(35,-5)*\cir(#1,0){}
,(40,-5)*\cir(#1,0){}
,(45,-5)*\cir(#1,0){}
,(5,-10)*\cir(#1,0){}
,(10,-10)*\cir(#1,0){}
,(15,-10)*\cir(#1,0){}
,(20,-10)*\cir(#1,0){}
,(25,-10)*\cir(#1,0){}
,(30,-10)*\cir(#1,0){}
,(35,-10)*\cir(#1,0){}
,(40,-10)*\cir(#1,0){}
,(45,-10)*\cir(#1,0){}
,(5,-15)*\cir(#1,0){}
,(10,-15)*\cir(#1,0){}
,(15,-15)*\cir(#1,0){}
,(20,-15)*\cir(#1,0){}
,(25,-15)*\cir(#1,0){}
,(30,-15)*\cir(#1,0){}
,(35,-15)*\cir(#1,0){}
,(40,-15)*\cir(#1,0){}
,(45,-15)*\cir(#1,0){}
,(5,-20)*\cir(#1,0){}
,(10,-20)*\cir(#1,0){}
,(15,-20)*\cir(#1,0){}
,(20,-20)*\cir(#1,0){}
,(25,-20)*\cir(#1,0){}
,(30,-20)*\cir(#1,0){}
,(35,-20)*\cir(#1,0){}
,(40,-20)*\cir(#1,0){}
,(45,-20)*\cir(#1,0){}
,(5,-25)*\cir(#1,0){}
,(10,-25)*\cir(#1,0){}
,(15,-25)*\cir(#1,0){}
,(20,-25)*\cir(#1,0){}
,(25,-25)*\cir(#1,0){}
,(30,-25)*\cir(#1,0){}
,(35,-25)*\cir(#1,0){}
,(40,-25)*\cir(#1,0){}
,(45,-25)*\cir(#1,0){}
,(5,-30)*\cir(#1,0){}
,(10,-30)*\cir(#1,0){}
,(15,-30)*\cir(#1,0){}
,(20,-30)*\cir(#1,0){}
,(25,-30)*\cir(#1,0){}
,(30,-30)*\cir(#1,0){}
,(35,-30)*\cir(#1,0){}
,(40,-30)*\cir(#1,0){}
,(45,-30)*\cir(#1,0){}
,(5,-35)*\cir(#1,0){}
,(10,-35)*\cir(#1,0){}
,(15,-35)*\cir(#1,0){}
,(20,-35)*\cir(#1,0){}
,(25,-35)*\cir(#1,0){}
,(30,-35)*\cir(#1,0){}
,(35,-35)*\cir(#1,0){}
,(40,-35)*\cir(#1,0){}
,(45,-35)*\cir(#1,0){}
,(5,-40)*\cir(#1,0){}
,(10,-40)*\cir(#1,0){}
,(15,-40)*\cir(#1,0){}
,(20,-40)*\cir(#1,0){}
,(25,-40)*\cir(#1,0){}
,(30,-40)*\cir(#1,0){}
,(35,-40)*\cir(#1,0){}
,(40,-40)*\cir(#1,0){}
,(45,-40)*\cir(#1,0){}
,(5,-45)*\cir(#1,0){}
,(10,-45)*\cir(#1,0){}
,(15,-45)*\cir(#1,0){}
,(20,-45)*\cir(#1,0){}
,(25,-45)*\cir(#1,0){}
,(30,-45)*\cir(#1,0){}
,(35,-45)*\cir(#1,0){}
,(40,-45)*\cir(#1,0){}
,(45,-45)*\cir(#1,0){}
#2\end{xy}}}

\newcommand{\rxyn}[2]{{\begin{xy} 0;<2mm,0mm>:<0mm,2mm>::0;0,
,(5,-2)*{a}
,(10,-1.8)*{b}
,(15,-2)*{c}
,(20,-2)*{d}
,(25,-2)*{e}
,(30,-2)*{f}
,(2,-5)*{a}
,(2,-10)*{b}
,(2,-15)*{c}
,(2,-20)*{d}
,(2,-25)*{e}
,(2,-30)*{f}
,(5,-5)*\cir(#1,0){}
,(10,-5)*\cir(#1,0){}
,(15,-5)*\cir(#1,0){}
,(20,-5)*\cir(#1,0){}
,(25,-5)*\cir(#1,0){}
,(30,-5)*\cir(#1,0){}
,(5,-10)*\cir(#1,0){}
,(10,-10)*\cir(#1,0){}
,(15,-10)*\cir(#1,0){}
,(20,-10)*\cir(#1,0){}
,(25,-10)*\cir(#1,0){}
,(30,-10)*\cir(#1,0){}
,(5,-15)*\cir(#1,0){}
,(10,-15)*\cir(#1,0){}
,(15,-15)*\cir(#1,0){}
,(20,-15)*\cir(#1,0){}
,(25,-15)*\cir(#1,0){}
,(30,-15)*\cir(#1,0){}
,(5,-20)*\cir(#1,0){}
,(10,-20)*\cir(#1,0){}
,(15,-20)*\cir(#1,0){}
,(20,-20)*\cir(#1,0){}
,(25,-20)*\cir(#1,0){}
,(30,-20)*\cir(#1,0){}
,(5,-25)*\cir(#1,0){}
,(10,-25)*\cir(#1,0){}
,(15,-25)*\cir(#1,0){}
,(20,-25)*\cir(#1,0){}
,(25,-25)*\cir(#1,0){}
,(30,-25)*\cir(#1,0){}
,(5,-30)*\cir(#1,0){}
,(10,-30)*\cir(#1,0){}
,(15,-30)*\cir(#1,0){}
,(20,-30)*\cir(#1,0){}
,(25,-30)*\cir(#1,0){}
,(30,-30)*\cir(#1,0){}
#2\end{xy}}}

\begin{document}

\thispagestyle{empty}

\begin{center}
CENTRE DE PHYSIQUE TH\'EORIQUE \footnote{\, Unit\'e Mixed de
Recherche (UMR) 6207 du CNRS et des Universit\'es Aix-Marseille 1 et 2 \\ \indent \quad \, Sud Toulon-Var, Laboratoire affili\'e \`a la 
FRUMAM (FR 2291)} \\ CNRS--Luminy, Case 907\\ 13288 Marseille Cedex 9\\
FRANCE
\end{center}

\vspace{1.5cm}

\begin{center}
{\Large\textbf{Almost-Commutative Geometries \\ Beyond the Standard Model II:
\\[2mm] New Colours}} 
\end{center}

\vspace{1.5cm}

\begin{center}
{\large Christoph A. Stephan
\footnote{\, stephan@cpt.univ-mrs.fr} }

\vspace{1.5cm}

{\large\textbf{Abstract}}
\end{center}

We will present an extension of the standard model of particle physics 
in its almost-commutative formulation. This extension is guided by
the minimal approach to almost-commutative geometries employed
in \cite{class}, although the model presented here is not
minimal itself. 

The corresponding almost-commutative geometry  leads 
to a Yang-Mills-Higgs model which consists of the standard model 
and two new fermions of opposite electro-magnetic charge
which may possess a new colour like gauge group. 
As a new phenomenon, grand unification is no longer required by
the  spectral action. 

\vspace{1.5cm}

\vskip 1truecm
\indent
CPT-P28-2007 \\
PACS-92: 11.15 Gauge field theories\\
MSC-91: 81T13 Yang-Mills and other gauge theories

\vskip 1truecm

\noindent March 2007

\newpage

\section{Introduction}

Understanding the origin of the standard model is currently one of
most challenging issues in high energy physics. Indeed, despite its
experimental successes, it is fair to say that its structure remains
a mystery. Moreover, a better understanding of its structure would
provide us with a precious clue towards its possible extensions.
This can be achieved in the framework of noncommutative geometry
\cite{book}, which is a branch of mathematics pioneered by Alain Connes,
 aiming at a generalisation of geometrical ideas to spaces whose
coordinates fail to commute. Motivated by quantum gravity, it is
postulated that space-time is a wildly noncommutative manifold at a
very high energy. Even if the precise nature of this noncommutative
manifold remains unknown, it seems legitimate to assume that at an
intermediate scale, say a few orders of magnitude below the Planck
scale, the corresponding algebra of coordinates is only a mildly
noncommutative algebra of matrix valued functions, called
almost-commutative geometries. When suitably
chosen, such a matrix algebra, a sum of three simple matrix algebras, 
reproduces within the spectral action
principle the standard model coupled to gravity \cite{cc}. 

Ten years after its discovery \cite{cc}, the spectral action has  recently received new impetus \cite{c06,barr,mc2} by allowing a  Lorentzian signature in the internal space. This mild modification  has three consequences.
The fermion-doubling problem \cite{2f} is solved elegantly,
Majorana masses and consequently the popular seesaw mechanism are  allowed for.
The Majorana masses in turn decouple the Planck mass from the $W$ mass.
Furthermore, Chamseddine, Connes \& Marcolli point out an additional  constraint on the coupling constants tying the sum of all Yukawa  couplings squared to the weak gauge coupling squared. This relation  already holds for Euclidean internal spaces
\cite{thum}.

For many years it has been tried  to construct models from  noncommutative geometry that go beyond the standard model \cite {beyond}. But these attempts failed to come up with anything physical if it was  not to add more generations and right-handed neutrinos to the  standard model.
For example the noncommutative constraints on the continuous parameters of the  standard model with four 
generations fail to be compatible with the  hypothesis of the big desert \cite{knecht}.

The situation changed recently, when a classification of  the finite
part of almost-commutative geometries with up to four summands in
the matrix algebra was performed \cite{class}.
This classification necessitated the heavy use of a computer program \cite{prog} 
to list the  irreducible Krajewski diagrams. Here the standard model appears
in a most prominent position. But also the so called electro-strong model 
was discovered which inspired the first viable almost-commutative model 
beyond the standard model: the $AC$-model \cite{chris}.
It comes from an algebra with six  summands  and is identical to the standard model with two  additional leptons $A^{--}$ and $C^{++}$ whose electric charge is two  in units of the electron charge. These new leptons  couple neither to  the charged gauge bosons, nor to the Higgs scalar. Their hypercharges  are vector-like, so that they do not contribute to the electroweak  gauge anomalies. Their masses are gauge-invariant and they constitute  viable candidates for cold dark matter \cite{klop}. 

In this paper
we will use a version of the standard model based on a matrix
algebra with four summands \cite{class}.
We will investigate a new extension of the standard model
which is also inspired by the classification
of irreducible almost-commutative geometries \cite{class} and
extends the standard model by $N$ generations of left-handed $SU(2)$
doublets and right-handed singlets. These new particles may also 
possess a new colour group, $SU(D)$, with respect to which the standard
model particles are singlets, i.e. neutral. They resemble closely to
the $\theta$-particles which were proposed by Lev Borisovich Okun
\cite{okun}.

One of the main  results of this paper is the fact  that the constraints on 
the gauge couplings of 
the standard model no longer resemble those
of grand unified theories. The new relation is given in equation (\ref{cutoff}). 
If extensions
of the standard model interact via the weak or the strong interactions,
this seems to be a general feature in almost-commutative geometry.
For colour groups $SU(D)$ with $D\geq 3$ one finds also that at least
three generations of new particles are needed.

The paper is organised as follows: We first give the basic notions
of a spectral triple, the main building  block of noncommutative geometry.
Then we quickly review how the Yang-Mills-Higgs model is obtained
via the fluctuated Dirac operator and the spectral action. This 
account is far from exhaustive and we refer to \cite{cc,farewell,mc2} for
a detailed presentation.

For the new particles the details of the spectral triple and the lift of 
the automorphisms are given.The Lagrangian as well as the 
constraints on the couplings are calculated. With help of the
one-loop renormalisation group equations the masses of the
new particles, the Higgs boson mass and, if applicable, the
value of the gauge coupling at low energies for the new colour 
group are calculated.

\section{Finite Spectral Triples}

In this section we will give the necessary basic definitions  of almost commutative
geometries from a particle physics point of view. 
For our calculations, only the finite part matters, 
so we restrict ourselves to real, finite spectral triples in $KO$-dimension
six:
($\mathcal{A},\mathcal{H},\mathcal{D}, $ $J,\chi$). Note that in the literature
before \cite{c06,barr,mc2} the finite part of the spectral triple was
considered to be of $KO$-dimension zero. The change in this algebraic dimension
amounts in some  sign changes, i.e. the commutator for the real structure
and the chirality changes into an anti-commutator and the anti-particles 
have opposite chirality with respect to the particles.

\subsection{Basic Definitions}

The algebra $\mathcal{A}$ is
a finite sum of matrix algebras
$\mathcal{A}= \oplus_{i=1}^{N} M_{n_i}(\mathbb{K}_i)$ with $\mathbb{K}_i=\mathbb{R},\mathbb{C},\mathbb{H}$ where $\mathbb{H}$
denotes the quaternions. 
A faithful representation $\rho$ of $\mathcal{A}$ is given on the finite dimensional Hilbert space $\mathcal{H}$.
The Dirac operator $\mathcal{D}$ is a selfadjoint operator on $\mathcal{H}$ and plays the role of the fermionic mass matrix.
$J$ is an antiunitary involution, $J^2=1$, and is interpreted as the charge conjugation
operator of particle physics.
The chirality $\chi$  is a unitary involution, $\chi^2=1$, whose eigenstates with eigenvalue
$+1$ $(-1)$ are interpreted as right (left) particle states and  $-1$ $(+1)$ for 
right (left) antiparticle states.
These operators are required to fulfill Connes' axioms for spectral triples:

\begin{itemize}
\item  $[J,\mathcal{D}]=\{J,\chi\}=\{ \mathcal{D},\chi\} =0$, 
 
$[\chi,\rho(a)]=[\rho(a),J\rho(b)J^{-1}]=
[[\mathcal{D},\rho(a)],J\rho(b)J^{-1}]=0, \forall a,b \in \mathcal{A}$.
\item The chirality can be written as a finite sum $\chi =\sum_i\rho(a_i)J\rho(b_i)J^{-1}.$
This condition is called {\it orientability}.
\item The intersection form
$\cap_{ij}:=\T(\chi \,\rho (p_i) J \rho (p_j) J^{-1})$ is non-degenerate,
$\rm{det}\,\cap\not=0$. The
$p_i$ are minimal rank projections in $\mathcal{A}$. This condition is called
{\it Poincar\'e duality}.
\end{itemize} 
Now the Hilbert space $\mathcal{H}$ and the representation are $\rho$ decomposed 
into left and right, particle and antiparticle spinors and representations:
\begin{eqnarray}
\mathcal{H}=\mathcal{H}_L\oplus\mathcal{H}_R\oplus\mathcal{H}_L^c\oplus\mathcal{H}_R^c \quad \quad 
\rho = \rho_L \oplus \rho_R \oplus \overline{ \rho_L^c} \oplus \overline{ \rho_R^c}
\notag
\label{representation}
\end{eqnarray}
In this representation the Dirac operator has the form
\begin{eqnarray}
\mathcal{D}=\pp{0&\mathcal{M}&0&0\\
\mathcal{M}^*&0&0&0\\ 0&0&0&\overline{\mathcal{M}}\\
0&0&\overline{\mathcal{M}^*}&0}, \label{opdirac}
\notag
\end{eqnarray}
where $\mathcal{M}$ is the fermionic mass matrix connecting the left and the right handed fermions.

Since the individual matrix algebras have only one fundamental representation for $\mathbb{K}=
\mathbb{R},\mathbb{H}$ and two for $\mathbb{K}=\mathbb{C}$ (the fundamental one and its complex
conjugate), $\rho$ may be written as a direct sum of these fundamental representations with
mulitiplicities
\begin{eqnarray}
\rho(\oplus_{i=1}^N a_i):=(\oplus_{i,j=1}^N
a_i \otimes
1_{m_{ji}} \otimes 1_{(n_j)})\
\oplus\ ( \oplus_{i,j=1}^N 1_{(n_i)} \otimes 1_{m_{ji}} \otimes
\overline{a_j} ).
\nonumber
\end{eqnarray}
The first summand denotes the particle sector and the second the antiparticle sector. For the dimensions
of the unity matrices we have $(n)=n$ for $\mathbb{K}=\mathbb{R},\mathbb{C}$ and $(n)=2n$ for
$\mathbb{K}=\mathbb{H}$ and the convention $1_0=0$.
The multiplicities $m_{ji}$ are non-negative integers. Acting with the real structure
$J$ on $\rho$ permutes the main summands and complex conjugates them. It is also possible to write
the chirality as a direct sum
\begin{eqnarray}
\chi=(\oplus_{i,j=1}^N 1_{(n_i)} \otimes \chi_{ji}1_{m_{ji}} \otimes
1_{(n_j)})\  
\oplus\ (\oplus_{i,j=1}^N 1_{(n_i)} \otimes (-\chi_{ji})1_{m_{ji}} \otimes 1_{(n_j)}),
\nonumber
\end{eqnarray}
where $\chi_{ji}=\pm 1$ according to the previous convention on left-(right-)handed spinors.
One can now define the multiplicity matrix $\mu \in M_N(\mathbb{Z})$ such that
$\mu _{ji}:=\chi _ {ji}\, m_{ji}$. This matrix is symmetric and decomposes into a particle and an antiparticle matrix, the second being just the particle matrix transposed, $\mu= \mu_P + \mu_A = \mu_P - \mu_P^T$. The intersection form of the Poincar\'e duality is now simply $\cap = \mu - \mu^T$, see \cite{kps}. Note that in contrast to
the case of $KO$-dimension zero, the multiplicity matrix is now antisymmetric rather
then symmetric.

\subsection{Obtaining the Yang-Mills-Higgs theory}

To construct the actual Yang-Mills-Higgs theory one starts out with the fixed (for convenience flat) Dirac operator of a 4-dimensional spacetime with a fixed fermionic
mass matrix. To generate curvature a general coordinate transformation is performed and then the Dirac operator is fluctuated. This can be achieved by lifting the automorphisms of the algebra to the Hilbert space, unitarily transforming the Dirac operator with the lifted automorphisms and then building linear combinations. Again it
is sufficient to restrict the treatment to the finite case.

All the automorphisms of matrix algebras connected to the
unity element, Aut$(\mathcal{A})^e$, are inner, i.e. they are of the form
\bb
i_u a = u a u^\ast \quad a\in \mathcal{A},
\ee
where
\bb
u \in \mathcal{U}(\mathcal{A}) = \{u\in \mathcal{A} | u^\ast u = u u^\ast = 1 \}
\ee
is an element of the group of unitaries of the algebra and $i$ is a map from the unitaries into the
inner automorphisms Int$(\mathcal{A})$
\bb
i : \mathcal{U}(\mathcal{A}) &\longrightarrow& \mbox{Int}(\mathcal{A}) \nonumber \\
u &\longmapsto& i_u.
\ee
In the kernel of $i$ are the central unitaries, which commute with all elements in 
$\mathcal{A}$.
These inner automorphisms Int$(\mathcal{A})$ are equivalent to the group of unitaries $\mathcal{U}(\mathcal{A})$ modulo the central unitaries $\mathcal{U}^c(\mathcal{A})$.

The Abelian algebras $\mathbb{R}$ and $\mathbb{C}$ do not possess any inner automorphisms. Remarkably the
quaternions and the matrix algebras over the complex numbers produce the kind of inner automorphisms
that correspond to the nonabelian gauge groups of the standard model. Note that the exceptional groups do
not appear. They are the automorphism groups of non-associative algebras.

As in the Riemannian case the automorphisms close to the identity are going to
be lifted to the Hilbert space. This lift has a simple closed form \cite{real}, $L=\hat{L}\circ i^{-1}$ with
\bb
\hat{L} (u) = \rho(u)J \rho(u) J^{-1}.
\ee
Here two crucial problems occur. The symmetry group of the standard model
contains an Abelian sub-group $U(1)_Y$. But the inner automorphisms do not contain any Abelian sub-groups by
definition.  Furthermore the lift is multivalued for matrix algebras over the complex numbers since the kernel of $i$ contains an $U(1)$-group. Note that neither the matrix algebras over the reals nor those over the quaternions have any central unitaries close to the identity. The solution to both of these problems is to centrally extend the lift, i.e. to adjoin some central elements  \cite{farewell}. One has to distinguish between central unitaries stemming from the Abelian algebra $\mathbb{C}$ and those from nonabelian matrix algebras $M_n(\mathbb{C})$, $n\geq 2$. To simplify let the algebra $\mathcal{A}$ be a sum of matrix algebras over the complex
numbers.  Furthermore the commutative and noncommutative sub-algebras will be separated,
\bb
\mathcal{A} = \mathbb{C}^M  \oplus \bigoplus_{k=1}^N M_{n_k}(\mathbb{C}) \ni (b_1,...,b_M,c_1,...,c_N), \quad
n_k \geq 2.
\ee
The group of unitaries $\mathcal{U}(\mathcal{A})$ and the group of central unitaries $\mathcal{U}^c(\mathcal{A})$ are then given by:
\bb
\mathcal{U}(\mathcal{A}_f)&=& U(1)^M \times U(n_1) \times ... \times U(n_N) \ni u=(v_1,...,v_M,w_1,...,w_N), \nonumber \\
\mathcal{U}^c(\mathcal{A}_f)&=& U(1)^{M+N} \ni u^c = (v_1,...,v_M,w_1^c  1_{n_1},..., w_N^c 1_{n_N}).
\ee
For the inner automorphisms follows
\bb
\mbox{Int}(\mathcal{A})=\mathcal{U}(\mathcal{A})/\mathcal{U}^c(\mathcal{A}) \ni u^{in}=(1,...,1,w^{in}_1, ...,w^{in}_N),
\ee
with $w^{in}_j \in \mathcal{U}(M_{n_j})/U(1)$. The lift $L=\hat{L}\circ i^{-1}$ can be written explicitely with
\bb
\hat{L} = \rho (1,...,1,w_1,...,w_M) J \rho(...) J^{-1}.
\ee
It is multivalued due to the kernel of $i$, ker$(i)=\mathcal{U}^c(\mathcal{A}_f)$. This multivaluedness can
be cured by introducing an additional lift $\ell$ for the central unitaries, which is restricted to those unitaries $\mathcal{U}^{nc}(\mathcal{A})$ stemming from 
the noncommutative part of the algebra,
\bb
\ell (w_1^c,...,w_N^c) &:=& \rho \left( \prod_{j_1=1}^N (w^c_{j_1})^{q_{1,j_1}} ,..., \prod_{j_M=1}^N (w^c_{j_M})^{q_{M,j_M}}, \prod_{j_{M+1}=1}^N (w^c_{j_{M+1}})^{q_{1,j_{M+1}}} 1_{n_1},... \right. \nonumber \\
&...,& \left. \prod_{j_{M+N}=1}^N (w^c_{j_{M+N}})^{q_{1,j_{M+N}}} 1_{n_N} \right)J\rho(...) J^{-1},
\ee
with the $(M+N)\times N$ matrix of charges $q_{k,j}$. The extended lift $\mathbb{L}$ is then defined
as
\bb
\mathbb{L} (u^i,w^c) := (\hat{L}\circ i^{-1})(u^i) \ell(w^c), \quad u^i\in \mbox{Int}
(\mathcal{A}), \; \; w^c \in \mathcal{U}^{nc}(\mathcal{A}).
\label{centralextendlift}
\ee
For convenience this lift will be written $\mathbb{L}(u)$ without making the
specific distinction between the unitaries and the central unitaries.

In this way Abelian gauge groups have been introduced and the multivaluedness has been reduced, depending on the choice of the matrix of charges.

The {\it fluctuation $\ddf$} of the Dirac operator $\mathcal{D}$ is given by a
finite collection $f$ of real numbers
$r_j$ and algebra automorphisms $u _j\in{\rm Aut}(\mathcal{A})^e$ such
that
\bb
\ddf :=\sum_j r_j\,\mathbb{L}(u _j) \, \mathcal{D} \, \mathbb{L}(u_j)^{-1},\quad r_j\in\mathbb{R}, u_j\in{\rm Aut}(\mathcal{A})^e.
\eee
These fluctuated Dirac operators build an affine space which serves as
the configuration space for the Yang-Mills-Higgs theory.
Only fluctuations with real coefficients are considered since $\ddf$ must remain selfadjoint. 
The sub-matrix of the fluctuated Dirac operator $\ddf$ which is equivalent to
the mass matrix $\mathcal{M}$,  is often denoted by $\varphi $, the
`Higgs scalar', in physics literature.

As mentioned in the introduction an almost commutative geometry is the tensor product of a finite
noncommutative triple with an infinite, commutative spectral triple. By
Connes' reconstruction theorem \cite{grav,av} it is known that the latter comes
from a Riemannian spin manifold, which will be taken to be any
4-dimensional, compact manifold.  The spectral
action of this almost-commutative spectral triple is
defined to be the number of eigenvalues of the Dirac operator $\ddf$ up to a cut-off $
\Lambda$. 
Via the heat-kernel expansion one finds, after a long and laborious calculation \cite{cc,mc2},
a Yang-Mills-Higgs action combined with the Einstein-Hilbert action, a cosmological
constant, a term containing the Weyl tensor squared  as well as a conformal 
coupling of the Higgs field to the curvature scalar:
\bb
S_{CC}[e,A_{L/R},\varphi] &=& \mbox{tr} \left[ h \left( \frac{\ddf^2}{\Lambda^2} \right) \right] 
\nonumber \\ \nonumber \\
&=& \int_M \left\{ \frac{2 \Lambda_c}{16 \pi G} - \frac{1}{16 \pi G} R + a (5R^2 - 8 R_{\mu\nu}
R^{\mu\nu} -7 R_{\mu\nu\lambda\tau}R^{\mu\nu\lambda\tau}) \right. \nonumber \\
&& + \sum_i \frac{1}{2 g_i^2} \mbox{tr}\ F^{\ast}_{i \mu \nu} F_i^{ \mu \nu} + \frac{1}{2} 
(D_\mu \varphi)^{\ast} D^\mu \varphi \nonumber \\
&& +\lambda \mbox{tr} (\varphi^{\ast} \varphi)^2 - \frac{1}{2} \mu^2 \mbox{tr} (\varphi^{\ast} \varphi)
\nonumber \\
&& + \left. \frac{1}{12} \mbox{tr} (\varphi^{\ast} \varphi) R \frac{}{} \right\} \; \rm{d}V + \mathcal{O}(\Lambda^{-2})
\label{CCaction}
\ee
where $h:\mathbb{R}_+ \rightarrow \mathbb{R}_+$ is a positive test function.
The coupling constants are functions of the  first moments $h_0$, $h_2$ and $h_4$ of
the test function
\bb
&\Lambda_c&=\alpha_1 \frac{h_0}{h_2} \Lambda^2, \; G= \alpha_2 \frac{1}{h_2} \Lambda^{-2}, \;
a = \alpha_3 h_4 , \nonumber \\
&g_i^2& = \alpha_{4i} \frac{1}{h_4}, \; \lambda = \alpha_5 \frac{1}{h_4}, \; \mu^2 = 
\alpha_5 \frac{h_2}{h_4} \Lambda^2.
\label{CCcouplings}
\ee
The curvature terms $F_{\mu \nu}$ and the covariant derivative $D_\mu$ are in the standard form 
of Yang-Mills-Higgs theory.
The constants $\alpha_j$ depend in general on the special choice of matrix
algebra and on the Hilbert space,
i.e. on the particle content. For details of the computation we refer to
\cite{cc,mc2}. 

This action is valid at the cut-off $\Lambda$ where it ties together the coupling constants $g_i$  of 
the gauge connections and the Higgs coupling $\lambda$ since they originate from the
same heat-kernel coefficient. 
For the standard model with three generations the calculation of the gauge couplings 
in (\ref{CCcouplings})
imposes at $\Lambda$ conditions on the $U(1)_Y$, $SU(2)_w$ and $SU(3)_c$
couplings $g_1$, $g_2$ and $g_3$ comparable to those of grand unified theories:
\bb
5 \, g_1^2 = 3 \, g_2^2 = 3 \, g_3^2 
\ee
But since the lift of the automorphisms produces extra free parameters through the
$U(1)$ central charges the first equality can always be fulfilled by a different 
choice of the central-charge. Therefore only the gauge couplings of
noncommutative gauge groups underlie constraints from the spectral action.

In the same way as for the gauge couplings the spectral action also implies
constraints for the quartic Higgs coupling $\lambda$ and the Yukawa
couplings. The full set of constraints for the standard model reads \cite{c06,
mc2,thum}:
\bb 
 3 \, g_2^2=  3 \, g_3^2= 3 \,\frac{Y_2^2}{H} \,\frac{\lambda}{24}\,= \,\frac{3}{4}\,Y_2\,.
\label{4con}
\ee
Here $Y_2$ is the sum of all Yukawa couplings $g_f$ squared, $H$ is  the sum of all Yukawa couplings raised to the fourth power. Our  normalisations are: $m_f=\sqrt{2}\,(g_f/g_2)\,m_W,$ $(1/2)\,(\partial  \varphi)^2+(\lambda/24)\,\varphi^4$.

As we will see in the following, the grand unified constraint $g_2^2 = g_3^2$ at the
cut-off $\Lambda$ is a very special case. It is valid for  the standard model
but in general it will not hold. The model presented in this paper is one
example for different constraints for $g_2$ and $g_3$ at the cut-off energy.
For possible extensions of the standard model within the framework of 
almost-commutative geometry, these constraints may limit the particle
content in a crucial way. 

\section{The spectral triple}

The basic entity of the spectral triple is the matrix algebra. For the 
model under consideration it is
\bb
\mathcal{A}= \mathbb{H} \oplus \mathbb{C}  \oplus M_3(\mathbb{C}) \oplus
\mathbb{C} \oplus M_D (\mathbb{C}) \oplus \mathbb{C} \ni (a,b,c,d,e,f).
\ee
It has the algebra $\mathcal{A}_{SM}$ of the standard model as a sub-algebra
and contains two new summands $\mathcal{A}_{new}$.
\bb
\mathcal{A}_{SM}= \mathbb{H} \oplus \mathbb{C}   \oplus M_3(\mathbb{C}) 
 \oplus \mathbb{C} \ \ {\rm and} \ \ \mathcal{A}_{new} =  M_D (\mathbb{C}) 
 \oplus \mathbb{C}
\ee
This particular model is inspired by the classification of almost-commutative
geometries presented in \cite{class}. For one generation in the standard model
and the new particles it is a slightly modified irreducible spectral
triple in the sense that one right-handed new particle can be deleted from the
spectral triple without violating the axioms. But in this case the physical model
would not be free of anomalies and has
therefore to be discarded. We do not want to go into all the details of the
construction, but for the interested reader we give the Krajewski diagram
of this almost-commutative spectral triple in the appendix.

The representation of the algebra on the Hilbert space is the usual one
for the standard model. For the new part the representation is given by
\bb
\rho_L (a)=  a \otimes 1_D,
\;
\rho_R (f)=  \pp{ f 1_D & 0  \\ 0 & \bar{f} 1_D }, 
\;
\rho_L^c (e)=    1_2 \otimes e,
\;
\rho_R^c (e)=  \pp{ e& 0 \\ 0 & e} .
\ee
The complete representation is then the direct sum of the standard model 
representation and the new part:
\bb
\rho = \rho_{SM} \oplus \rho_{new} \ \  {\rm with} \  \ \rho_{new}(a,e,f) = \rho_L (a) \oplus \rho_R (f) \oplus \rho_L^c (e) \oplus \rho_R^c (e)
\ee
The same holds for the Hilbert space, $\mathcal{H} = \mathcal{H}_{SM} \oplus \mathcal{H}_{new}$. For one generation of new particles the Hilbert space is
\bb
\mathcal{H}_{new} = ( \mathbb{C}^2 \otimes \mathbb{C}^D)
\oplus (\mathbb{C} \otimes \mathbb{C}^D) \oplus (\mathbb{C} \otimes \mathbb{C}^D) \oplus \ {\rm antiparticles}.
\ee
The dimension of $\mathcal{H}_{new}$ depends on the number of generations $N$ of the 
new particles and the size $D$ of the sub-algebra $M_D(\mathbb{C})$ and
reads dim$(\mathcal{H}_{new})= 8\, N \, D$.

We will denote the spinors of the new particles by $\psi_1$ and $\psi_2$. They
appear as left-handed $SU(2)$-doublets and right-handed $SU(2)$-singlets
\bb
\pp{\psi_{1} \\ \psi_{2}}_L \oplus (\psi_{1})_R \oplus (\psi_{2})_R \oplus \pp{\psi_{1}^c \\ \psi_{2}^c}_L \oplus 
(\psi_{1}^c)_R \oplus (\psi_{2}^c)_R \in \mathcal{H}_{new}.
\ee
Furthermore every $\psi_i$ is a $SU(D)$ D-plet for $D \geq 2$. The Dirac operator 
contains the Yukawa couplings of the new particles
\bb
\mathcal{M}_{new}= \pp{g_{\psi_1}&0 \\ 0&g_{\psi_2} } \otimes 1_D \ \ {\rm with}
\ \ g_{\psi_1},g_{\psi_2} \in \mathbb{C}.
\ee
and is given by
\bb
D_{new} = \pp{ 0 & \mathcal{M}_{new} & 0 & 0 \\ \mathcal{M}_{new}^* &0&0&0 \\
0&0&0& \overline{\mathcal{M}}_{new} \\ 0&0&\overline{\mathcal{M}}_{new}^{\,*}&0}.
\ee
For more than one generation it is off course possible to introduce a CKM-type
matrix which mixes the generations. But to keep the analysis of the model as
simple as possible we will not include family mixing. 

From the Krajewski diagram, figure \ref{kra1} in the appendix, 
it is straightforward to see that
all the axioms for the spectral triple are fulfilled. To the three generations of 
the standard model one may add any number $N$ of generations of the
new particles with an arbitrarily large sub-algebra $M_D(\mathbb{C})$.
In the following we will investigate the physical models with respect
to the number of generations of new particles $N$ and with respect
to the  size $D$ of the sub-algebra.

\section{The gauge group, the lift and the constraints}

The group of unitaries of the noncommutative part of the algebra is
\bb
\mathcal{U}^{nc}(\mathcal{A})= SU(2)_w \times U(3) \times U(D) \ni (v,w,s) \ \
{\rm for } \ \ D \geq 2.
\ee
In the case of $D=1$ the group is just $\mathcal{U}^{nc}(\mathcal{A})= SU(2) \times 
U(3)$, as for the standard model. 
Define $u:= \det (w) \in U(1)_1$ and $r := \det (s) \in U(1)_2$. Note that in
case of $D=1$, $s$ and $r$ can
simply be dropped from the following calculations. 

The lift of the unitaries decomposes into a standard model part and a part
for the new particles,
\bb
\mathbb{L} ( v, u^{p_1} s^{q_1},u^{p_2} s^{q_2} w ,u^{p_3} s^{q_3},u^{p_4} s^{q_4} r,
u^{p_5} s^{q_5})=&&\mathbb{L}_{SM} ( v, u^{p_1} s^{q_1},u^{p_2} s^{q_2} w ,u^{p_3} s^{q_3}) \nonumber \\ 
 &\oplus &  \mathbb{L}_{new} (v, u^{p_4} s^{q_4} r, u^{p_5} s^{q_5})
\ee
with $p_i,q_i \in \mathbb{Z}$. This lift produces a priori two $U(1)$ groups 
through the central extensions $u$ and $r$.
But it  has been shown in \cite{class} that in the case of two $U(1)$ 
groups the requirement of being anomaly free results in proportional couplings
of the $U(1)$'s to the
standard model particles. The two photons can therefore be linearly combined into a physical
photon and an unphysical photon that does not couple to the standard model.
Without loss of generality we can therefore set $q_1=q_2=q_3=0$. For the
standard model part of the lift one finds then  $p_1=-p_3=-1/2$ and
$p_2=1/6 - 1/3$ from anomaly cancellation. This reduces $U(3)$ to 
$U(1)_Y \times SU(3)_c$ in the correct representation. 

The exact form of the lift for the new particles is given by
\bb
\mathbb{L}_{new} (v, u^{p_4} s^{q_4} r, u^{p_5} s^{q_5}) =
{\rm diag}[ u^{p_4} s^{q_4} v \otimes r ; u^{p_4+p_5} s^{q_4+q_5} r, 
u^{p_4-p_5} s^{q_4-q_5} r].
\ee
Being anomaly free and requiring that corresponding left-handed and  right-handed
particles are equally charged under the little group leads to $p_4=q_5=0$,
$q_4= -1/D$ and $p_5= p_1 = -1/2$. 
Since the normalisation of the lift for the right-handed electron is chosen 
to be $Y_{e_R} = 2 p_1 = -1$, one
sees immediately that $Y_{\psi_L} = 0$, $Y_{\psi_R} = \pm p_1 = \mp 1/2$.
Therefore the electro-magnetic charge of the new particles is $Q_{el}=\pm 1/2 e $.
This charge assignment is summarised in the following table:
\begin{center}
\begin{tabular}{|c|c|c|c|c|c|}
\hline
&I & $I_3$ & $Y_{new}$ & $Q_{el}$ & $SU(D)$ \\ 
\hline &&&&& \\
$(\psi_{1})_L$ & $2$ & $+\frac{1}{2}$ & 0 & $+\frac{e}{2}$ & $D$ \\
&&&&& \\
\hline
&&&&& \\
$(\psi_{2})_L$ & $2$ & $-\frac{1}{2}$ & 0 & $-\frac{e}{2}$ & $D$ \\
&&&&& \\
\hline
&&&&& \\
$(\psi_{1})_R$ & 1 &  0  & $\frac{1}{2}$ & $+\frac{e}{2}$ & $D$ \\
&&&&& \\
\hline
&&&&& \\
$(\psi_{2})_R$ & 1 & 0 & $-\frac{1}{2}$ & $-\frac{e}{2}$ & $D$ \\
&&&&& \\
\hline
\end{tabular}
\end{center}
Plugging in the numbers one finds for lift
\bb
\mathbb{L}_{new} (v, s^{-1/D} r, u^{-1/2} ) &=&
{\rm diag}[  s^{-1/D} v \otimes r ; u^{-1/2} s^{-1/D} r, 
u^{1/2} s^{-1/D} r] \nonumber \\
&=& {\rm diag}[  v \otimes \tilde{r} ; u^{-1/2} \tilde{r}, 
u^{1/2}\tilde{r}] 
\ee
with $\tilde{r} \in SU(D)_{new}$. For $D \geq 2$ the gauge group of the
model is then
\bb
U(1)_Y \times SU(2)_w \times SU(3)_c \times SU(D)_{new}
\ee
and for $D=1$ it is just the standard model gauge group.
It is also remarkable that the inner fluctuations of
the mass matrix $\mathcal{M}_{new}$ with the lift $\mathbb{L}_{new}$
produce exactly the standard model Higgs field
\bb
\sum_j r_j\,\mathbb{L}_{L,new}(v_i,  \tilde{r}_i, u_i^{-1/2} ) \, \mathcal{M}_{new} \, \mathbb{L}^{-1}_{R,new}(v_i,  \tilde{r}_i, u_i^{-1/2} )
= \varphi_{SM} \, \mathcal{M}_{new} ,
\ee
where the subscripts $L$ and $R$ indicate the left-handed and the right-handed
parts of  the lift. Therefore $\mathcal{M}_{new}$ contains the Yukawa couplings of the
new model, in exact analogy to the standard model.

From the spectral action one obtains now immediately the Lagrangian for
the new particles,
\bb
\mathcal{L}_{new} =&
-&\frac{1}{4} \rm{tr}(G_{\mu\nu}G^{\mu\nu})
+  i \sum_{i=1..N} (\bar{\psi_1},\bar{\psi_2})^i_L D^\psi_L \pp{\psi_1 \\ \psi_2}^i_L 
 \nonumber \\
&+&  i \sum_{i=1..N} (\bar{\psi_1})^i_R D^{\psi_1}_R (\psi_1)^i_R+ i \sum_{i=1..N} (\bar{\psi_2})^i_R D^{\psi_2}_R (\psi_2)^i_R 
\nonumber \\
& -& \sum_{i=1..N} (g_{\psi_1})_i (\bar{\psi_1},\bar{\psi_2})^i_L \varphi_{SM} (\psi_1)^i_R 
 - \sum_{i=1..N} (g_{\psi_2})_i (\bar{\psi_1},\bar{\psi_2})^i_L \varphi_{SM} (\psi_2)^i_R 
\nonumber \\ 
 &+& \mbox{hermitian conjugate},
\ee
where the covariant derivatives are given by
\bb
D^{\psi}_L &=& \partial_\mu+ i g_2 W^k_\mu \frac{\tau_k}{2} +i g_4 G^a_\mu t_a  \\
D^{\psi_i}_R &=& \partial_\mu + i g_1 \frac{Y_i}{2} B_\mu +i g_4 G^a_\mu t_a. 
\ee
Here $g_1$ and $g_2$ are the standard model $U(1)_Y$ and $SU(2)_w$ gauge 
couplings. For $D\geq 2$  the $SU(D)$ gauge coupling is $g_4$, $t_a$ are
the corresponding generators and $G^a_\mu$ are the  gauge fields
with the usual curvature tensor
\bb
G_{\mu \nu} &=& \partial_\mu G_\nu - \partial_\nu G_\mu - g_4 [G_\mu,G_\nu]. 
\ee
The $SU(D)$ terms have off course to be dropped from all equations
if $D=1$.

From the spectral action it is now straight forward to calculate the constraints
on the gauge couplings, the quartic Higgs coupling and the  Yukawa couplings.
The normalisation of the quartic
Higgs coupling is taken to be the same as for the standard model:
\bb
N\, g_4^2 = 3 \, g_3^2=  \left(3+ \frac{N \, D}{4}\right) \, g_2^2= 3 \,\frac{Y_2^2}{H} \,\frac{\lambda}{24}\,= \,\frac{3}{4}\,Y_2\,.
\label{4con}
\ee
$Y_2$ and $H$ include now the Yukawa couplings of the new particles in
the standard way. 

One notes immediately that models beyond the standard
model in almost-commutative geometry will in general not exhibit 
the constraint $g_2=g_3$ from grand unified theories. This is a qualitatively
new feature and may prove to be important in restricting possible
extensions of the standard model within the frame work of the
spectral action.

In the following analysis we will for simplicity assume that all the  Yukawa
couplings of the new particles in all generations are equal, i.e.  
$g_{\psi_1} = g_{\psi_2} =: g_{\psi}$. 
For more realistic models one would 
off course admit different Yukawa couplings, but as a first estimation
of the particle masses equal couplings should be sufficient. Furthermore 
we will assume three generations for the standard model particles and
we will neglect all standard model Yukawa couplings safe the top quark
coupling $g_t$. 
Under these  assumptions the constraints on the couplings at the cut-off
$\Lambda$ read:
\bb
g_3^2 &=& \left(1+ \frac{N \, D}{12}\right) g_2^2 \label{cutoff} \\
g_4^2 &=& \left( \frac{3}{N} + \frac{D}{4} \right) g_2^2 \label{g4} \\
g_t^2 &=& \frac{4 + \frac{N \, D}{3}}{3+ 2 N \, D \, R^2} \, g_2^2 \ \ {\rm with}
\ \ R\,:= \, \frac{g_\psi}{g_t}  \label{gtop} \\
\lambda &=& 8 \, \left( 3 + \frac{N\, D}{4} \right) 
\frac{3 + 2 N \, D \, R^4 }{(3 + 2 N \, D \, R^2)^2 } \, g_2^2 \label{lambda}
\ee 
Here it becomes obvious that equation (\ref{cutoff}) no longer
resembles the grand unification condition $g_3^2 = g_2^2$ as in
the case of the pure standard model.

The strategy to find the masses of the new particles and the mass of 
the Higgs boson is now the following. With help of the renormalisation
group equation for $g_2$ and $g_3$ one determines the cut-off
via condition (\ref{cutoff}). There one can fix $g_4$ and $\lambda$ 
with conditions (\ref{g4}) and (\ref{lambda}). The last free parameter
is the ratio $R$ of  the
Yukawa coupling $g_\psi$ of the new particles and  the top quark 
Yukawa coupling $g_t$.
This ratio is fixed by the requirement that the renormalisation group
flow produces the measured pole mass of the top quark, $m_t=170.9\,\pm2.6$ GeV
\cite{data}.

\section{The renormalisation group equations.}

We will now give the one-loop $\beta$-functions of the standard model   with three generations of standard model particles, $N$ generations of the new particles
with either no new colour, i.e. $D=1$ and  $M_D(\mathbb{C}) = \mathbb{C}$,
or with an $SU(D)$ colour group and $D\geq 2$. They will serve to evolve the
constraints (\ref{4con}) from $E= \Lambda$  down to our energies $E=m_Z$. We set:
$ t:=\ln (E/m_Z),\qq \de g/\de t=:\beta _g,\qq \kappa :=(4\pi )^{-2}$.  
We will neglect all standard model fermion masses  below the top mass and also  
neglect threshold effects. 

The $\beta$-functions are \cite{mv,jones}:
\bb 
\beta _{g_i}&=&\kappa b_ig_i^3,\qq 
b_i=
{\textstyle
\left( \frac{41}{6}  + \frac{N \, D}{3},-\frac{19}{6}+\frac{N \, D}{3},
-7, - \frac{11}{3} N + \frac{4}{3} D  \right)  },
\\ \cr
\beta _t&=&\kappa
\left[ -\sum_i c_i^ug_i^2 +Y_2 +\,\frac{3}{2}\,g_t^2
\,\right] g_t,\\
\beta _{\psi_1}&=&\kappa
\left[ -\sum_i c_i^\psi g_i^2 +Y_2 +\,\frac{3}{2}\,g_{\psi_1}^2
-\,\frac{3}{2}\,g_{\psi_2}^2\, \right] g_{\psi_1},\\
\beta _{\psi_2}&=&\kappa
\left[ -\sum_i c_i^\psi g_i^2 +Y_2 +\,\frac{3}{2}\,g_{\psi_2}^2
-\,\frac{3}{2}\,g_{\psi_1}^2\, \right] g_{\psi_2},\\
\beta _\lambda &=&\kappa
\left[ \,\frac{9}{4}\,\left( g_1^4+2g_1^2g_2^2+3g_2^4\right)
-\left( 3g_1^2+9g_2^2\right) \lambda
+4Y_2\lambda -12H+4\lambda ^2\right] ,\ee
with
\bb c_i^t=\left( {\textstyle\frac{17}{12}},{\textstyle\frac{9}{4}} , 8 \right) ,
&
c_i^\psi =\left( {\textstyle\frac{3}{4}},{\textstyle\frac{9}{4}} , 0, 3 \, \frac{D^2 - 1}{D} \right) ,\\
Y_2=3g_t^2+ N \, D g_{\psi_1}^2 +N \, D g_{\psi_2}^2,
&
H=3g_t^4+N \, D g_{\psi_1}^4+N \, D g_{\psi_2}^4.
\ee
For the case $D=1$ the $\beta$-functions of $g_4$ and $c_4^\psi$ are to
be ignored.
The gauge couplings decouple from the other equations and have  
identical evolutions in both energy domains:
\bb 
g_i(t)=g_{i0}/\sqrt{1-2\kappa b_ig_{i0}^2t}.
\ee
The initial conditions are taken from experiment \cite{data}:
\bb 
g_{10}= 0.3575,\qq
g_{20}=0.6514,\qq 
g_{30}=1.221.
\label{ggg}
\ee
Then the unification  scale $\Lambda $ is the solution of 
$\left(1+ \frac{N\, D}{12} \right) \,g_2(\ln (\Lambda /m_Z))=
g_3(\ln  (\Lambda /m_Z))$,
\bb 
\Lambda = m_Z\exp\frac{g_{20}^{-2}-\left(1+ \frac{N\, D}{12} \right)^2 \, g_{30}^{-2}}{2\kappa (b_2-\left(1+ \frac{N\, D}{12} \right)^2 \,b_3)},
\ee
and  depends on  the number of generations
of new particles $N$ and on the size of the matrix algebra $D$. 

\section{The masses and the couplings at $m_Z$}

We require that all couplings remain  perturbative and
we obtain the pole masses of the Higgs, the top and  the new particles:
\bb 
m_H^2=\,\frac{4}{3}\,\frac{\lambda(m_H) }{g_2(m_Z)^2}\,m_W^2,\qq
m_t=\sqrt{2}\,\frac{g_t(m_t)}{g_2(m_t)}\,m_W,\qq
m_{\psi_1}=\,m_{\psi_2}=\,\frac{g_\psi(m_\psi)}{g_2(m_Z)^2}\,m_W^2.
\ee
As experimental input we have the initial conditions of the three 
standard model gauge couplings (\ref{ggg}) and the mass of the
top quark, $m_t=170.9\,\pm2.6$ GeV
\cite{data}. As mentioned before the masses of the  new particles are 
assumed to be equal. With the constraints (\ref{cutoff}) to (\ref{lambda})
we can now determine their numerical value via the renormalisation 
group flow and we can also determine the mass of the Higgs boson
for the respective model.

Let us start with the case $D=1$, $M_D(\mathbb{C})=\mathbb{C}$.
The gauge group for this model is just the standard model gauge 
group.
For up to three generations of new particles, the resulting masses
and cut-off energies are summarised in the following table:
\begin{center}
\begin{tabular}{|c|c|c|c|}
\hline
N  & $\Lambda$ [GeV] & $m_H$ [GeV] & $m_\psi$ [GeV] \\
\hline
1 & $5,3 \cdot 10^{13}$ & $167,3 \pm 3,4$ & $69,3 \mp 3,5$ \\  
2 & $3,0 \cdot 10^{11}$ & $172,0 \pm 3,2$ & $53,7 \mp 2,5$ \\
3 & $7,4 \cdot 10^9$ & $177,8 \pm 2,5$ & $48,0 \mp 1,6$ \\
\hline
\end{tabular}
\end{center}
The new particles are very light. And since they possess electro-magnetic
charge they should clearly have been detected if they existed \cite{data}.
This model has therefore to be discarded.

Next we consider the case $D=2$. The gauge group for the model
is $U(1)_Y \times SU(2)_w \times SU(3)_c \times SU(2)_{new}$. The standard
model particles and the Higgs boson are $SU(2)_{new}$ singlets.
For one generation it is not possible to solve the constraint
(\ref{g4}) within the real numbers, i.e. $g_{4}(t)$ has a pole below
the cut-off. For two and three generations the resulting masses
and cut-off energies are: 
\begin{center}
\begin{tabular}{|c|c|c|c|c|}
\hline
N  & $\Lambda$ [GeV] & $m_H$ [GeV] & $m_\psi$ [GeV] & $g_4(m_Z)$\\
\hline
2 & $4,4 \cdot 10^{8}$ & $182,3 \pm 2,3$ & $69,6 \mp 2,3$ & 1,50 \\
3 & $8,3 \cdot 10^6$ & $196,7 \pm 2,5$ & $50,2 \mp 1,7$ & 0,91 \\
\hline
\end{tabular}
\end{center}
The detectability of these particles is not as obvious as in the case
without a new colour. First of all, the gauge coupling $g_4$ is strong,
so one should expect confinement. Therefore the new particles will,
as quarks, not appear as free particles but bound into colour singlets.
These composite particles could allow to escape from detectors
if they are neutral and thus hide the new particles from detection.
It is beyond the scope of this paper to give an analysis of 
the phenomenological details of the models, so we will postpone
this analysis for a later publication.

For $D\geq 2$ the gauge coupling $g_{4}(t)$ has a pole below the cut-off for one 
and two generations of new particles. So three generations
is the minimal number. We give the cut-off energies, the masses of the  
new particles and the Higgs mass with respect to $D$ are for three
generations of new particles:
\begin{center}
\begin{tabular}{|c|c|c|c|c|}
\hline
D  & $\Lambda$ [GeV] & $m_H$ [GeV] & $m_\psi$ [GeV] & $g_4(m_Z)$\\
\hline
3 & $2,1 \cdot 10^{5}$ & $217,0 \pm 1,5$ & $56,5 \mp 1,1$ & 1,22 \\
4 & $2,0 \cdot 10^4$ & $241,2\pm 1,6$ & $62,3 \mp 1,3 $ & 1,54 \\
5 & $4,1 \cdot 10^3$ & $268,3 \pm 0,8$ & $65,8 \mp 0,7$ & 1,78 \\
6 & $1,3 \cdot 10^3$ & $300,6 \pm 0,8$ & $63,6 \mp 0,7$ & 1,81 \\
7 & $524 $ & $338,2 \pm 0,7$ & $57,6 \mp 0,4$ & 1,70 \\
8 & $261$ & $379,3 \pm 0,9$ & $50,7 \mp 0,5$ & 1,53 \\
\hline
\end{tabular}
\end{center}
With respect to the detectability the same arguments apply as 
for the case $D=2$.
The coupling $g_4$ is very strong for all models, so small
confinement radii are to be expected. 
It is also interesting to note that the mass of the  Higgs boson
is strongly dependend on the cut-off scale. This reminds of
 the older Connes-Lott model \cite{lott} which predicted 
 a Higgs mass of $m_{H,C-L} \sim 250 -  324 $GeV \cite{gut}.  
For $D\geq 8$ the cut-off
energy becomes smaller than the Higgs mass. This is also an
interesting new feature that deserves further investigation.

\section{Conclusions and outlook} 

We have presented a particle model based on an almost-commutative
geometry which contains the standard model as a sub-model. It provides
an extension of the standard model with $N$ generations of new
particles. These particles come as a  left-handed $SU(2)_w$ doublet
and  two right-handed singlets. The requirement of being anomaly free
forces them to have opposite electro-magnetical charges with an 
absolute value of half the electron charge. Furthermore the model
allows these particles to have a new $SU(D)_{new}$ colour. In this case
they are equivalent to Okun's $\theta$-particles \cite{okun}.

The spectral action puts strong constraints on the gauge couplings,
the quartic Higgs coupling  and the Yukawa couplings of this model
at the cut-off scale. Using standard renormalisation group equations
these constraints allow to calculate the masses of the new particles
and the mass of the Higgs boson at low energies. The masses of
the new particles, under the assumption of equal masses in all
families, range then from $\sim 48$ GeV to $\sim 69$ GeV, where
up to three generations and new colour groups up to $SU(8)_{new}$
have been considered.
For the Higgs boson the masses range from $ \sim 164$ GeV  to 
$\sim 334$ GeV. 

One should  note that these new particles could be the preferred the
decay products of the Higgs boson. If they can hide from direct
detection this would mean that the Higgs  boson could also be much more
difficult to detect.

It is also interesting to note that the cut-off scale of the spectral
action is considerably lowered by the presence of these new
particles. For three generation and eight colours it sinks even
below the Higgs mass, $\Lambda \sim 261$ GeV for a Higgs
mass of $m_H \sim 334$ GeV.
Remarkably for colour groups larger than $SU(2)_{new}$ one has 
to add at least three generation of the new particles. 

Many questions have not been considered in this article.
Le us list some of these: 

\begin{itemize}
\item Are the new particles directly detectable by existing experiments such
as LEP and Tevatron?
\item Will the new particles be detectable by future experiments such as LHC?
\item What is the phenomenology of the model?  What are  for example the
stable colour
singlet states and confinement radii?
\item Does the model contain a viable dark matter
candidate?
\end{itemize}

This list is certainly not exhaustive and other interesting 
questions may arise. But the model shows clearly how
difficult extension of the standard model within almost-commutative
geometry will be in general.
The constraints stemming from the spectral action together
with the geometrical constraints from the spectral triple
formalism restrict model building severely. Apart from
the older $AC$-model \cite{chris,klop}, which possesses
a viable dark matter candidate, this is so far the only
model which could be in concordance with experiment.

\vskip1cm
\noindent
{\bf Acknowledgements:} The author would like to thank T. Sch\"ucker,
Andr\'e Tilquin and
M. Y. Khlopov for helpful comments and discussions. We
gratefully acknowledge a fellowship of the Alexander von Humboldt-Stiftung.

\section*{Appendix: The Krajewski Diagram}

\begin{figure}
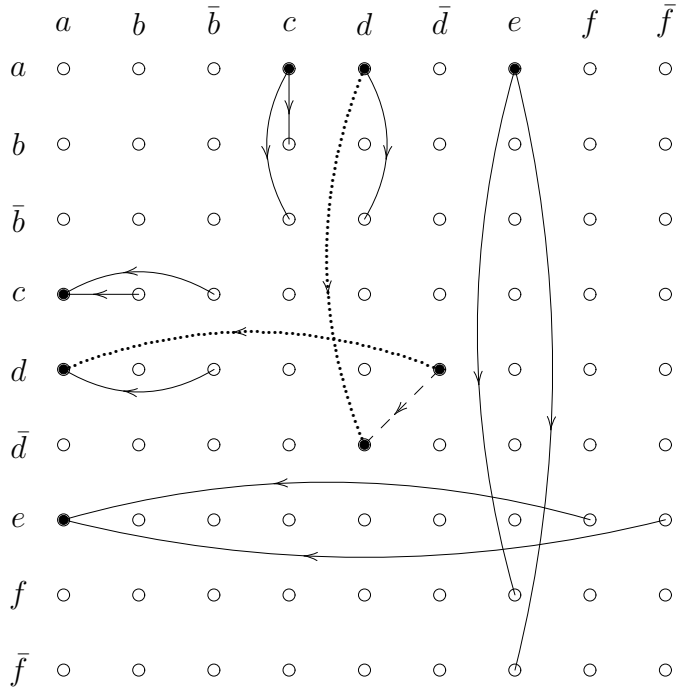

\begin{center}
\begin{tabular}{c}
\rxyz{0.4}{
,(5,-20)*\cir(0.3,0){}*\frm{*}
,(5,-25)*\cir(0.3,0){}*\frm{*}
,(5,-20);(10,-20)**\dir{-}?(.4)*\dir{<}
,(5,-20);(15,-20)**\crv{(10,-17)}?(.4)*\dir{<}
,(5,-25);(15,-25)**\crv{(10,-28)}?(.4)*\dir{<}
,(5,-25);(30,-25)**\crv{~*=<2pt>{.}(17.5,-20)}?(.45)*\dir{<}
,(20,-5)*\cir(0.3,0){}*\frm{*}
,(25,-5)*\cir(0.3,0){}*\frm{*}
,(20,-5);(20,-10)**\dir{-}?(.6)*\dir{>}
,(20,-5);(20,-15)**\crv{(17,-10)}?(.6)*\dir{>}
,(25,-5);(25,-15)**\crv{(28,-10)}?(.6)*\dir{>}
,(25,-5);(25,-30)**\crv{~*=<2pt>{.}(20,-17.5)}?(.6)*\dir{>}
,(30,-25)*\cir(0.3,0){}*\frm{*}
,(25,-30)*\cir(0.3,0){}*\frm{*}
,(25,-30);(30,-25)**\dir{--}?(.4)*\dir{<}
,(5,-35)*\cir(0.3,0){}*\frm{*}
,(5,-35);(40,-35)**\crv{(22.5,-30)}?(.4)*\dir{<}
,(5,-35);(45,-35)**\crv{(25,-40)}?(.4)*\dir{<}
,(35,-5)*\cir(0.3,0){}*\frm{*}
,(35,-5);(35,-40)**\crv{(30,-22.5)}?(.6)*\dir{>}
,(35,-5);(35,-45)**\crv{(40,-25)}?(.6)*\dir{>}
} 
\end{tabular}
\end{center}
\caption{ Krajewski diagram of the standard model with right-handed 
neutrino  and Majorana-mass term depicted by the dashed arrow. 
The new 
particles reside  in the $e$-line and  the $e$-column.
}
\label{kra1}
\end{figure}

In this appendix we present the Krajewski diagrams which were
used to construct the model treated in this publication. 
Krajewski diagrams do for spectral triples  what the Dynkin and 
weight diagrams do for groups and  representations.
For an introduction into the formalism of Krajewski we refer to \cite{kps,class}.
The Krajewski diagram for the model presented in this paper 
is depicted in figure \ref{kra1}. It shows one generation of standard model
particles and one generation of new particles.

The arrows encoding the new particles are drawn on the $e$-line
and  the $e$-column.
Note the similarity to the standard model quark
sector which sits on the $c$-line and  the $c$-column. The dotted 
arrows denote the possible right-handed neutrinos and the
dashed arrow represents  a possible Majorana mass term.

\begin{figure}
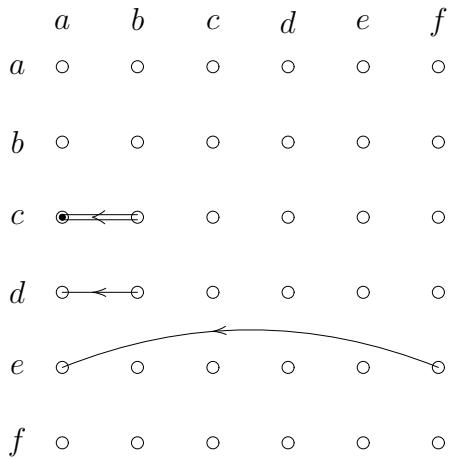

\begin{center}
\begin{tabular}{c}
\rxyn{0.4}{
,(5,-20);(10,-20)**\dir{-}?(.4)*\dir{<}
,(5,-15);(10,-15)**\dir2{-}?(.4)*\dir2{<}
,(5,-15)*\cir(0.2,0){}*\frm{*}
,(30,-25);(5,-25)**\crv{(17.5,-20)}?(.6)*\dir{>}
} \\
\\
\end{tabular}
\end{center}
\caption{Minimal Krajewski diagram associated with the diagram 
in figure \ref{kra1}.}
\label{kra2}
\end{figure}

This diagram originates from the minimal diagram shown in figure
\ref{kra2}. One remarks immediately that the right-handed neutrinos
as well as one of the right-handed new particles can be neglected
from the purely geometric point of view. But this model is not anomaly
free and has therefore to be excluded. 

The multiplicity matrix $\mu$ associated to the Krajewski diagram in figure \ref{kra1},
with three generations of standard model particles and $N$ generations
of new particles, can be directly read off to be
\bb
\mu = \pp{ 0&0&0&0&0&0 \\ 0&0&0&0&0&0 \\-3&6&0&0&0&0 \\-3&3&0&0&0&0
\\ - N&0&0&0&0& 2N \\0&0&0&0&0&0}
\ee
The axiom of the Poincar\'e duality is fulfilled since
$\det (\mu - \mu^t ) = 81 \, (2 N)^2 \neq 0 \ {\rm for \ all} \ N \in \mathbb{N}$.
Only the right-handed neutrinos violate the  axiom of orientability, 
\cite{ko6}, which is also the case for the pure standard model.

\end{document}